\documentclass[aps,prl,a4paper,twocolumn,superscriptaddress,citeautoscript,footinbib,floatfix]{revtex4-2}
\usepackage{color}
\usepackage[english]{babel}
\usepackage{lmodern}
\usepackage[T1]{fontenc}
\usepackage{amsmath}
\usepackage{units}
\usepackage{grffile}
\usepackage{color}
\usepackage[bottom]{footmisc}
\usepackage{booktabs}
\usepackage{dcolumn}
\usepackage{graphicx}
\usepackage{xcolor}
\usepackage{lipsum}
\usepackage[normalem]{ulem}
\graphicspath{{figs/}}
\begin{document}
\title{Role of $p$–$d$ Hybridization on Optical Properties of Chalcopyrite Semiconductors}
\author{Neunghee Han}\thanks{These authors contributed equally to this work.}
\affiliation{Department of Semiconductor and Chemical engineering,
Jeonbuk National University, Jeonju 54896, Republic of Korea}
\author{Harang Kim}\thanks{These authors contributed equally to this work.}
\affiliation{Department of Semiconductor and Chemical engineering,
Jeonbuk National University, Jeonju 54896, Republic of Korea}
\author{Minjae Kim}\email{mjkim1985@jbnu.ac.kr}
\affiliation{Department of Semiconductor Science and Technology,
Jeonbuk National University, Jeonju 54896, Republic of Korea}
\author{Woonhyuk Baek}\email{whbaek@jbnu.ac.kr}
\affiliation{Department of Semiconductor and Chemical engineering,
Jeonbuk National University, Jeonju 54896, Republic of Korea}
\date{\today}
\begin{abstract}
Designing quantum materials for coherent optical properties is a central agenda in quantum technology.
Semiconductor quantum dots are an emerging approach for controlling coherent optical properties via confinement effects, tunable band gaps, and exciton binding energies, yet their inherent structural and compositional inhomogeneity degrades the coherence of the optical spectra, posing a major obstacle.
We show that, for chalcopyrite semiconductors, hybridization between transition-metal $d$ and ligand $p$ electrons in the valence band is key to the coherence of the quantum dot optical spectrum. We demonstrate this using first-principles electronic-structure calculations and optical spectroscopy.
The strong $p$-$d$ hybridization in CuInS$_{2}$ induces the Cu($d$) Coulomb scattering channel, giving rise to the incoherent photodoped hole carrier, while the weak $p$-$d$ hybridization in AgInS$_{2}$ induces the delocalized photodoped hole carrier having a predominant S($p$) orbital character.
Our experimental results on optical spectra suggest that when the Cu ratio is enhanced in the Ag$_{1-x}$Cu$_{x}$In$_{1-y}$Ga$_{y}$S$_{2}$ quantum dot, Cu atoms at both Ag sites and defect sites
experience enhanced $p$-$d$ hybridization, and a coupling begins to develop between the electrons in the quantum dot and the defect electrons at a small Cu ratio.
This coupling activates Cu($d$) Coulomb scattering for photodoped holes traversing the defect sites, producing an incoherent optical response that naturally explains the long-standing absence of band-edge spectral signatures in CuIn$_{1-y}$Ga$_y$S$_2$ quantum dots.
These results serve as a guideline for designing semiconductor quantum dots. To achieve a coherent optical spectrum, avoid $p$-$d$-hybridized orbital character in the photo-doped carrier.
\end{abstract}
\maketitle

The manipulation of the coherent optical properties of semiconducting quantum materials is the essence of photonics technology~\cite{koch2006semiconductor}.
Quantum dots are a promising platform for this area of photonics technology,
by controlling the band gap and the exciton binding energy as a function of the size of the quantum dot~\cite{garcia2021semiconductor}.
Taking advantage of the high surface-to-bulk ratio, the quantum dots provide a building block to design nanostructures for photoluminescent, catalytic, and photovoltaic applications~\cite{baek2021highly,kim2015highly,kim2019moisture,park2025red,liu2026core,jiang2023development,sundstrom2026engineering,mann2025diminishing,motomura2023quantum}.

However, in many cases, quantum dot compounds exhibit structural and compositional inhomogeneity
with decreasing the distance from the surface,
inducing a decoherence in the optical spectrum~\cite{chae20263d,kim2025designing}.
To overcome this obstacle,
designing semiconductor quantum materials
with optical properties immune to structural and compositional inhomogeneity is required~\cite{kim2025designing}.

The I--III--VI$_2$ chalcopyrite semiconductor quantum dots have been suggested for heavy-metal free, eco-friendly photoluminescent and photovoltaic materials~\cite {kim2015highly,kim2019moisture,park2025red,liu2026core,jiang2023development,sundstrom2026engineering,mann2025diminishing,motomura2023quantum}.
Building on this, shelling the AgInS$_{2}$ quantum dot with a wide-gap semiconductor such as GaS$_{x}$, AgIn$_{1-y}$Ga$_{y}$S$_{2}$, enhances the coherence of the photoluminescent spectrum by eliminating the surface dangling bond effects~\cite{park2025red,liu2026core,jiang2023development,sundstrom2026engineering,mann2025diminishing,motomura2023quantum}.
Interestingly, a small amount of Cu doping for the quantum dot, Ag$_{1-x}$Cu$_{x}$In$_{1-y}$Ga$_{y}$S$_{2}$, brings a substantial red shift of the band edge photoluminescent spectra~\cite{liu2026core} followed by a complete suppression of the spectra upon enhancing the Cu ratio~\cite{jiang2023development}.
This Cu doping-induced evolution of the band edge spectra raises a question about the role of the Cu($d$) electrons in the presence of structural and compositional inhomogeneity.

In this Letter, we demonstrated that in I--III--VI$_2$ chalcopyrite semiconductors,
the $p$-$d$ hybridization between ligand and transition-metal atoms is a key factor for coherent optical properties
immune to disorder involving the transition-metal atoms.
The strong $p$-$d$ hybridization induces a Coulomb-scattering channel for the transition-metal atoms' $d$ orbitals.
This Coulomb scattering gives rise to the decoherence in the optical spectrum, including the exciton's spectrum.
This phenomenon is demonstrated with the computational results for CuInS$_{2}$ and AgInS$_{2}$ chalcopyrite semiconductor materials,
including Cu/Ag-In anti-site disorder and Cu/Ag deficiency induced defects to the bulk, and the experimental results
for the optical spectrum of Ag$_{1-x}$Cu$_{x}$In$_{1-y}$Ga$_{y}$S$_{2}$ quantum dots.

We performed first-principles calculations using the density-functional-theory plus Hubbard-$U$ plus intersite-$V$ (DFT+$U$+$V$) method
~\cite{PhysRevX.5.011006,lee2020first,tancogne2020parameter,jang2023intersite,hohenberg1964inhomogeneous,kohn1965self,leiria2010extended}
as implemented in the QUANTUM ESPRESSO package~\cite{giannozzi2009quantum}.
The on-site and intersite Hubbard interaction parameters, $U$ and $V$, were determined self-consistently using the extended ACBN0 ($e$ACBN0) method~\cite{PhysRevX.5.011006,lee2020first,tancogne2020parameter,jang2023intersite}.
Experimental crystal structures of CuInS$_2$ and AgInS$_2$ were adopted for the electronic-structure calculations~\cite{Spiess1974,Delgado2001}.
Computational details and experimental procedures for the synthesis and the characterization of Ag$_{1-x}$Cu$_x$In$_{1-y}$Ga$_y$S$_2$ (AIGS) quantum dots are provided in the Supplemental Material (SM)~\cite{suppl}.

\begin{figure}[t]
\includegraphics[width=\columnwidth]{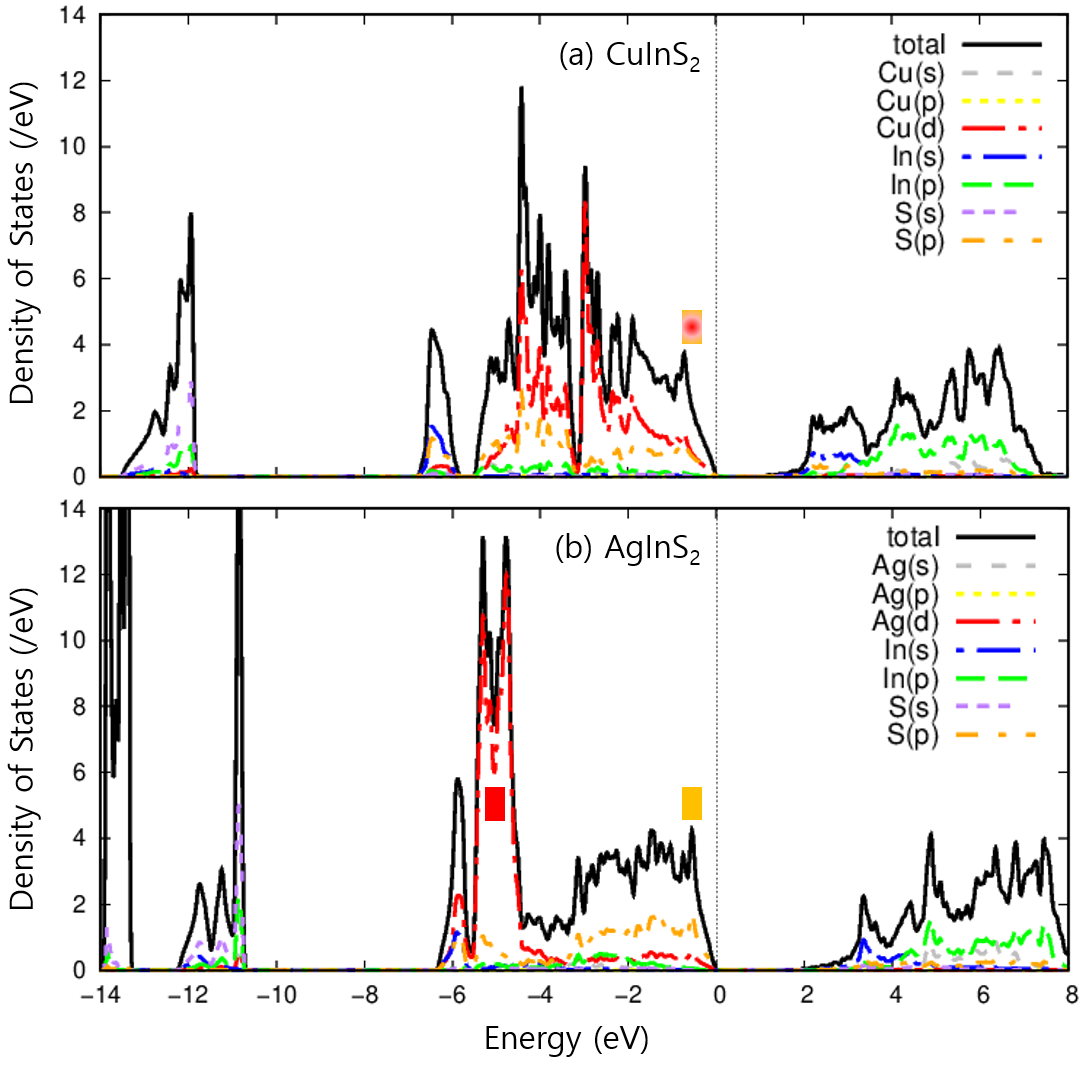}
\caption{(a) Density of states of CuInS$_{2}$ bulk.
The bar with a color gradient presents the
energy for the Cu($d$)-S($p$)
hybridized peak structure in the density of states.
(b) Density of states of AgInS$_{2}$ bulk.
The red and orange colored bars present
the energies for the
Ag($d$) and the S($p$) dominant density of states, respectively,
having a smaller $p$-$d$ hybridization
compared to the case of CuInS$_{2}$.
The total density of states are for the formula unit.
\label{figure:DOS_Pristine}
}
\end{figure}

Figure \ref{figure:DOS_Pristine} presents the density of states of pristine CuInS$_{2}$ and AgInS$_{2}$ bulk.
In the nominal ionic state, transition metal (TM), In, and S atoms have a valency of [TM]$^{1+}$In$^{3+}$S$^{2-}_{2}$.
For the $d^{10}$ configuration of the [TM]$^{1+}$ state, Cu($d$) electrons are strongly localized compared to the Ag($d$) electrons,
inheriting the smaller ionization energy for the Cu($d$) electrons compared to the ionization energy for the Ag($d$) electrons.
As shown in Fig.\ref{figure:DOS_Pristine} (a and b), this factor results in
the Cu($d$) electrons' energy level closer to the S($p$) electrons' energy level in CuInS$_{2}$
while the Ag($d$) electrons' energy level is far below to the S($p$) electrons' energy level in AgInS$_{2}$.

The difference in the TM($d$) and S($p$) energy levels induces
a fundamental impact on the character of the valence band Bloch wave function.
As shown in Fig.\ref{figure:DOS_Pristine}(a),
CuInS$_{2}$ has a strong S($p$)-Cu($d$) hybridization due to its smaller energy difference,
revealing the $p$-$d$ hybridization peak in the density of states for the valence band.
This factor inherits S($p$)-Cu($d$) hybridized orbital character for the Bloch wave function of the band.
The photon induced hole carrier has multiplet characters of both $d^{9}$ and $d^{10}\underline{L}$,
involves many-body effects from the Cu($d$) electrons~\cite{hughes2018copper}.
On the other hand, as shown in Fig.\ref{figure:DOS_Pristine}(b), AgInS$_{2}$ have a weaker S($p$)-Ag($d$) hybridization
and Ag($d$) energy level far below the S($p$) energy level compared to the CuInS$_{2}$.
This factor inherits S($p$) dominant and weakly S($p$)-Ag($d$) hybridized orbital character
for the Bloch wave function of the valence band for AgInS$_{2}$.
The photon induced hole carrier has a multiplet character of predominant $d^{10}\underline{L}$,
involves a delocalization of the hole carrier~\cite{hughes2018copper}.

Fig.\ref{figure:DOS_Pristine} (a and b) presents that,
the orbital character of the conduction band has a dominant In($s$) orbital with a finite,
smaller contribution from the In($p$) orbital, in both CuInS$_{2}$ and AgInS$_{2}$.
This similar conduction band orbital character and the strongly (weakly) $p$-$d$ hybridized orbital character in CuInS$_{2}$ (AgInS$_{2}$)
explains the smaller band gap in CuInS$_{2}$ compared to AgInS$_{2}$.
The strong $p$-$d$ hybridization in CuInS$_{2}$ induces a repulsion between the bonding and antibonding orbitals,
giving rise to an upward shift in the valence band energy level and reducing the band gap.

\begin{table}[b]
\caption{On-site Hubbard ($U$) and inter-site extended Hubbard ($V$) parameters for CuInS$_2$ and AgInS$_2$ bulks
within the $e$ACBN0 framework~\cite{PhysRevX.5.011006,lee2020first,tancogne2020parameter}.}
\label{tab:interaction_compare}
\begin{ruledtabular}
\begin{tabular}{lccc}
 & \multicolumn{1}{c}{CuInS$_2$ (eV)} &  & \multicolumn{1}{c}{AgInS$_2$ (eV)} \\
Interaction & Parameter & Interaction & Parameter \\
\hline
$U(\mathrm{Cu}\text{-}d)$ & 5.12 & $U(\mathrm{Ag}\text{-}d)$ & 6.68 \\
$U(\mathrm{S}\text{-}p)$  & 1.53 & $U(\mathrm{S}\text{-}p)$  & 2.63 \\
$U(\mathrm{In}\text{-}p)$ & 0.02 & $U(\mathrm{In}\text{-}p)$ & 0.04 \\
$V(\mathrm{Cu}\text{-}d,\, \mathrm{S}\text{-}p)$ & 2.36 & $V(\mathrm{Ag}\text{-}d,\, \mathrm{S}\text{-}p)$ & 2.13 \\
$V(\mathrm{Cu}\text{-}d,\, \mathrm{S}\text{-}s)$ & 1.03 & $V(\mathrm{Ag}\text{-}d,\, \mathrm{S}\text{-}s)$ & 1.17 \\
$V(\mathrm{In}\text{-}p,\, \mathrm{S}\text{-}s)$ & 0.44 & $V(\mathrm{In}\text{-}p,\, \mathrm{S}\text{-}s)$ & 0.51 \\
$V(\mathrm{In}\text{-}p,\, \mathrm{S}\text{-}p)$ & 0.53 & $V(\mathrm{In}\text{-}p,\, \mathrm{S}\text{-}p)$ & 0.87 \\
$V(\mathrm{In}\text{-}s,\, \mathrm{S}\text{-}s)$ & 0.60 & $V(\mathrm{In}\text{-}s,\, \mathrm{S}\text{-}s)$ & 0.50 \\
$V(\mathrm{In}\text{-}s,\, \mathrm{S}\text{-}p)$ & 0.86 & $V(\mathrm{In}\text{-}s,\, \mathrm{S}\text{-}p)$ & 0.79 \\
\end{tabular}
\end{ruledtabular}
\end{table}

Table \ref{tab:interaction_compare} presents the on-site Hubbard ($U$)
and inter-site extended Hubbard ($V$) interaction parameters of CuInS$_{2}$ and AgInS$_{2}$
within the $e$ACBN0 framework~\cite{PhysRevX.5.011006,lee2020first,tancogne2020parameter} (See SM~\cite{suppl}).
Noteworthy is that the on-site Hubbard interaction, $U$, for the Cu($d$) and S($p$) orbital
in CuInS$_{2}$ is smaller than the Ag($d$) and S($p$) orbital in AgInS$_{2}$,
while the inter-site extended Hubbard interaction, $V$, for the TM($d$)-S($p$) orbitals in CuInS$_{2}$ is larger than that in the AgInS$_{2}$.
These smaller $U$ and larger $V$ values in TM($d$)-S($p$) orbitals in CuInS$_{2}$ compared to AgInS$_{2}$ imply that
the occupied Bloch wave functions' charge has larger (smaller) component in TM-S bonding orbital (TM, S atomic orbital)
in CuInS$_{2}$ compared to AgInS$_{2}$, consistent with the computational results in Fig.\ref{figure:DOS_Pristine}.

\begin{figure}[t]
\includegraphics[width=\columnwidth]{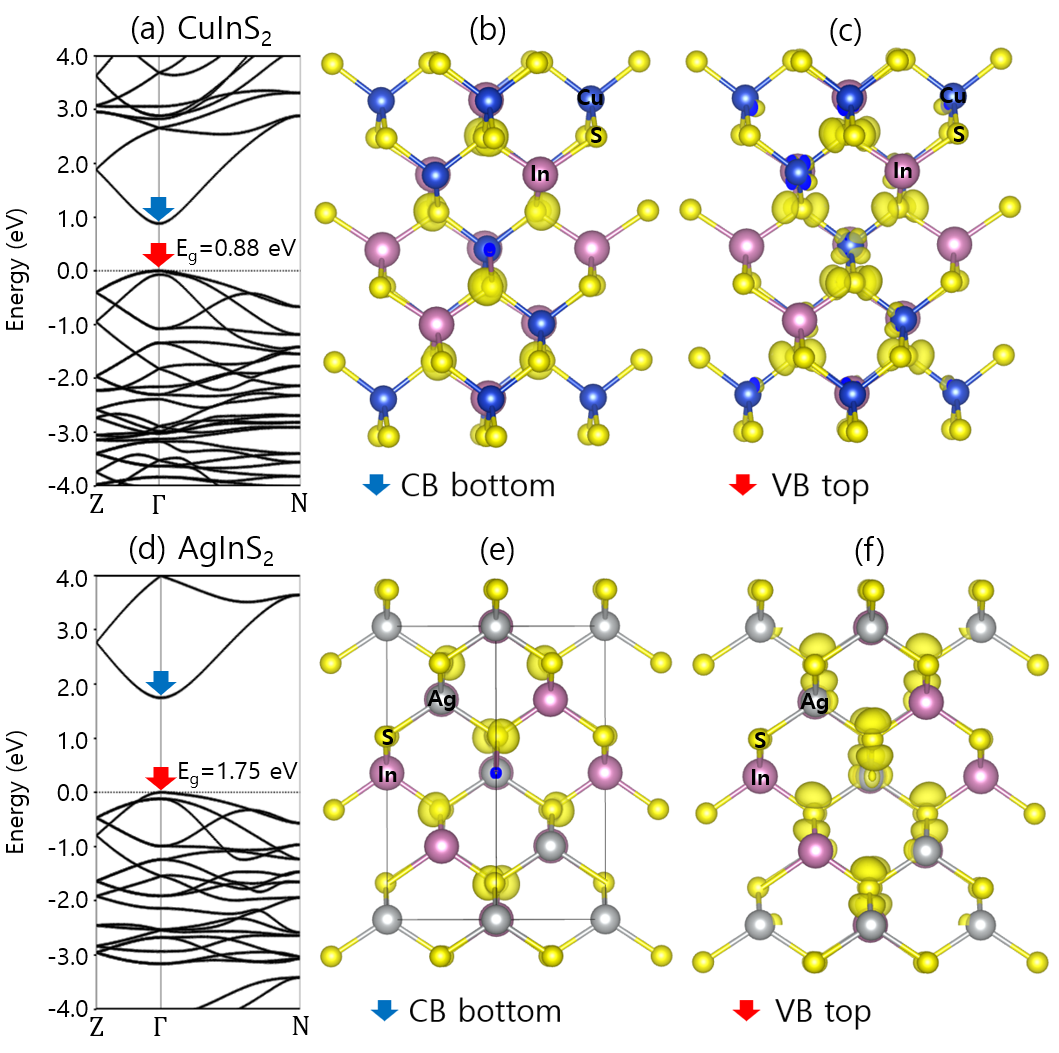}
\caption{(a) The band structure of CuInS$_{2}$ bulk.
The direct band gap (E$_{g}$) at $k=\Gamma$ is 0.88 eV.
(b, c) The charge densities
for the wave function of
(b) the conduction band (CB) bottom and
(c) the valence band (VB) top
at $k=\Gamma$ are presented,
indicated by blue and red arrows in (a), respectively.
(d) The band structure of AgInS$_{2}$ bulk.
The direct band gap (E$_{g}$) at $k=\Gamma$ is 1.75 eV.
(e, f) The charge densities
for the wave function of
(e) the conduction band (CB) bottom and
(f) the valence band (VB) top
at $k=\Gamma$ are presented,
indicated by blue and red arrows in (d), respectively.
The contour for the charge densities
is set to 0.002 $e$/$Å^{3}$.
\label{figure:Band_Charge_Pristine}
}
\end{figure}

Figure \ref{figure:Band_Charge_Pristine} presents the band structures and the charge densities
of the valence-band top (VBT) and the conduction-band bottom (CBB) at $k=\Gamma$.
The overall band structure in Fig.\ref{figure:Band_Charge_Pristine}(a and d),
the lighter electron band
and the heavier hole band in both CuInS$_{2}$ and AgInS$_{2}$,
is consistent with the literature~\cite{shabaev2015energy,jaffe1983electronic,liu2015electronic}. The band gap in AgInS$_{2}$ in the present result is 1.75 eV,
close to the experimental band gap of 1.87 eV~\cite{aguilera2007photoluminescence}, comparable in accuracy with hybridization functions (1.93 eV)~\cite{liu2015electronic},
improved from the DFT results of 0.27 eV.
The band gap in CuInS$_{2}$ in the present result is 0.88 eV,
closer to the experimental band gap of 1.55 eV~\cite{tell1971electrical}, and shows a comparable improvement compared to the previous results using a hybridization function, 1.24 eV~\cite{shabaev2015energy},
concerning the metallic electronic structures in the DFT.

The VBT and CBB charge densities at $k=\Gamma$  in Fig.\ref{figure:Band_Charge_Pristine}(b, c, e, and f)
reveals the difference of the S($p$)-TM($d$) hybridization between CuInS$_{2}$ and AgInS$_{2}$.
The Bloch wave functions of VBT and CBB are responsible for the optical excitation from the direct band gap,
and forms the exciton state when the interaction-driven kernel for the electron-hole binding is included.
Interestingly, consistent with the density of states in Fig.\ref{figure:DOS_Pristine},
the VBT charge density in CuInS$_{2}$ reveals the strongly S($p$)-Cu($d$) hybridized orbital character.
As shown in Fig.\ref{figure:Band_Charge_Pristine}(c),
the lobes of Cu($d$) (S($p$)) wave functions are directed toward nearest-neighboring S (Cu) sites.
On the other hand, as shown in Fig.\ref{figure:Band_Charge_Pristine}(f),
the VBT wave function of AgInS$_{2}$ reveals the weakly S($p$)-Ag($d$) hybridized orbital character.
The lobes of S($p$) wave functions are directed toward the $c$ axis, while the lobes of Ag($d$) wave functions are in the $ab$ plane.

As shown in Fig.\ref{figure:Band_Charge_Pristine}(b and e),
the CBB orbital characters in CuInS$_{2}$ and AgInS$_{2}$ are
similar to each other, consistent with the density of states in Fig.\ref{figure:DOS_Pristine}.
Given that the VBT charge densities are directed toward bond direction in CuInS$_{2}$
and crystalline axis in AgInS$_{2}$, the overlap of CBB and VBT wave functions
is stronger for CuInS$_{2}$ than that for AgInS$_{2}$.
This $p$-$d$ hybridization-driven difference in the CBB and VBT wave functions
implies that the TM($d$) electron highly impacts the coherence of the exciton spectrum.

\begin{figure}[t]
\includegraphics[width=\columnwidth]{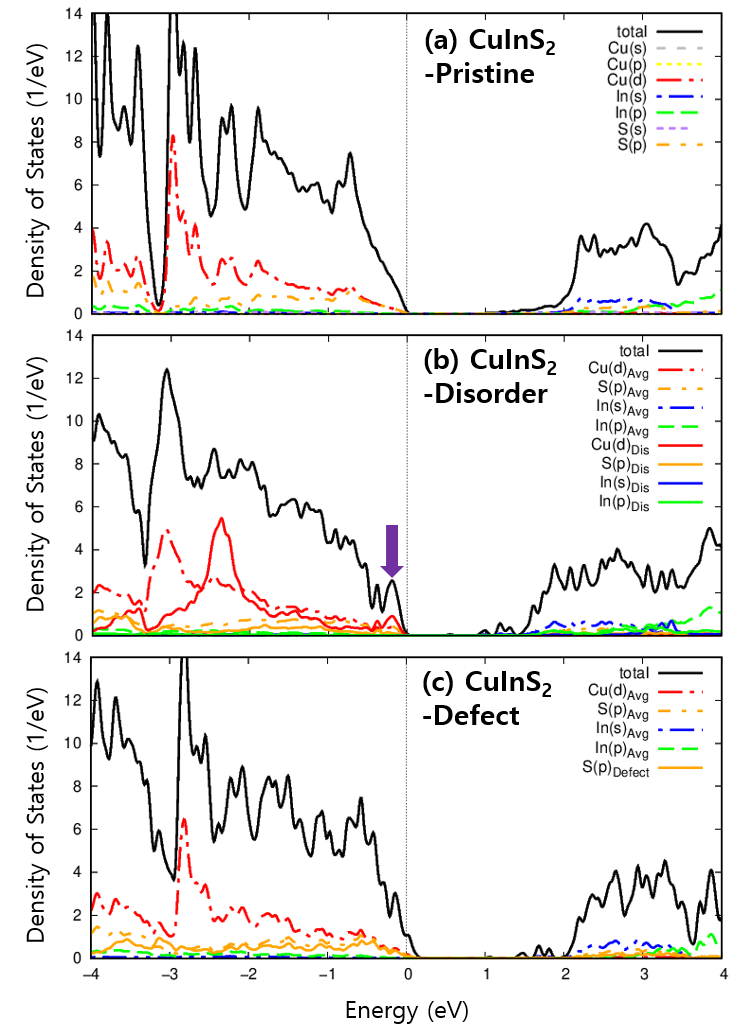}
\caption{(a, b, c) The electronic structures of CuInS$_{2}$ bulk
for (a) the pristine structure, (b) the anti-site disorder
of Cu-In, and (c) the defect from the Cu deficiency.
See the main text for the details of the structures.
Averaged values over the supercell structure are presented by Avg.
The disordered Cu and In sites, and the nearest neighboring S sites
are presented by Dis.
The nearest neighboring S sites from the Cu defect site
is presented by Defect.
The total density of states are for the unit cell of the pristine
structure.
The purple arrow in (b) presents
the anti-site disorder driven defect states at the Cu site,
strongly hybridized with its environment atoms' orbital.
\label{figure:DOS_Disorder_Defect_CuInS2}
}
\end{figure}

Figure \ref{figure:DOS_Disorder_Defect_CuInS2} presents
comparisons between pristine CuInS$_{2}$ bulk and
CuInS$_{2}$ with Cu-In anti-stie disorder or Cu deficiency.
The $2\times2\times1$ supercell is used for the
calculations with disorder and Cu defect (See SM~\cite{suppl}).
The nearest-neighboring Cu-In sites are considered
for the anti-site disorder.
This anti-site disorder and Cu defect
could be inherited by the spatial inhomogeneity in the quantum dot.

\begin{figure}[t]
\includegraphics[width=\columnwidth]{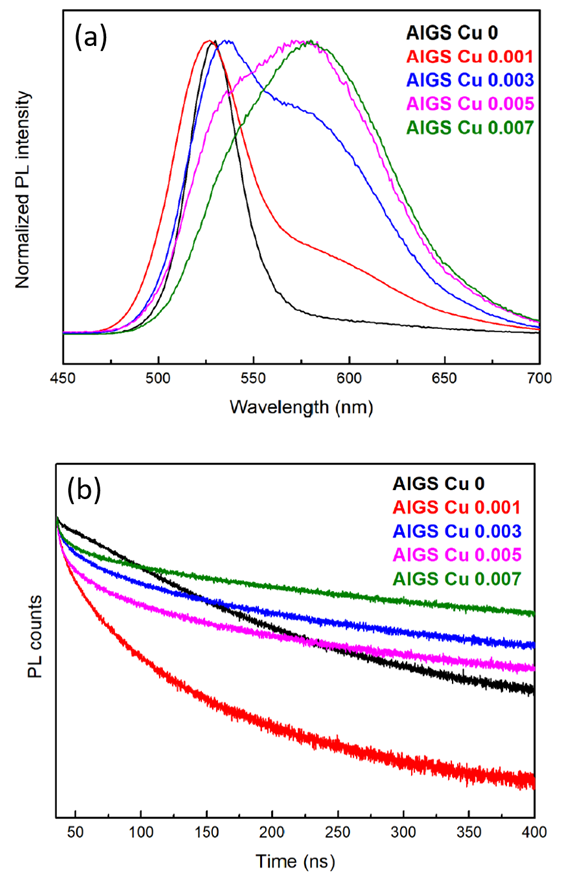}
\caption{(a) Band-edge photoluminescence (PL) spectra of
Ag$_{1-x}$Cu$_{x}$In$_{1-y}$Ga$_{y}$S$_2$ (AIGS) as a function of Cu composition.
(b) Time-resolved photoluminescence decay of
AIGS as a function of Cu composition.
\label{figure:PL_AIGS}
}
\end{figure}

From the anti-site disorder, a strong electronic dipole forms upon switching between the Cu$^{+}$ and In$^{3+}$ ions,
inducing a discontinuity in the electrostatic potential.
As shown in Fig.\ref{figure:DOS_Disorder_Defect_CuInS2}(a and b),
due to the strong S($p$)-Cu($d$) hybridization,
the discontinuity of the electrostatic potential is screened by the S($p$) electron.
The defect band is broadened by the S($p$)-Cu($d$) hybridization.
We suggest that this defect-environment hybridization is the origin of the incoherence in the optical conductivity
in the Ag$_{1-x}$Cu$_{x}$In$_{1-y}$Ga$_{y}$S$_{2}$ quantum dot
with enhanced Cu ratio $x$, as shown in our experimental results in Figure~\ref{figure:PL_AIGS}.
The strong electronic Coulomb scattering between the anti-site Cu($d$) electron
and the VB electron gives rise to an incoherent optical spectrum.
On the other hand, the Cu deficiency induces a relatively weak electrostatic defect from the absence of Cu$^{+}$ ion,
weakly impacts the electronic structure as a hole doping effect compared to the pristine case,
as shown in Fig.\ref{figure:DOS_Disorder_Defect_CuInS2}(a and c).

\begin{figure}[t]
\includegraphics[width=\columnwidth]{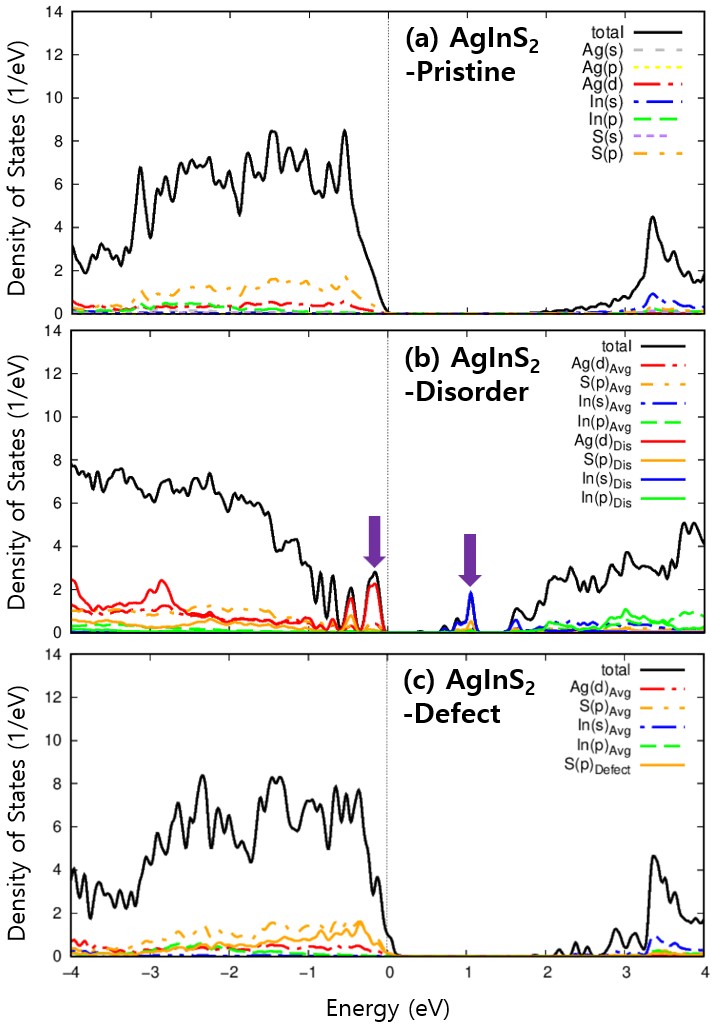}
\caption{(a, b, c) The electronic structures of AgInS$_{2}$ bulk
for (a) the pristine structure, (b) the anti-site disorder
of Ag-In, and (c) the defect from the Ag deficiency.
See the main text for the details of the structures.
Averaged values over the supercell structure are presented by Avg.
The disordered Ag and In sites, and the nearest neighboring S sites
are presented by Dis.
The nearest neighboring S sites from the Ag defect site
is presented by Defect.
The total density of states are for the unit cell of the pristine
structure.
The purple arrows in (b) present
the anti-site disorder driven defect states at the
Cu and In sites,
strongly localized and disconnected from environment atoms' orbital.
\label{figure:DOS_Disorder_Defect_AgInS2}
}
\end{figure}

\begin{table}[b]
\caption{On-site Hubbard interaction parameters $U$ for Cu($d$) and Ag($d$) orbitals in different structural configurations
within the $e$ACBN0 framework~\cite{PhysRevX.5.011006,lee2020first,tancogne2020parameter}.}
\label{tab:U_values}
\begin{ruledtabular}
\begin{tabular}{lcc}
Structure & $U(\mathrm{Cu}\text{-}d)$ (eV) & $U(\mathrm{Ag}\text{-}d)$ (eV) \\
\hline
Pristine        & 5.12        & 6.68        \\
Disordered site & 5.58        & 1.24        \\
TM Defects      & 4.45--4.75  & 6.27--6.45  \\
\end{tabular}
\end{ruledtabular}
\end{table}

Figure \ref{figure:DOS_Disorder_Defect_AgInS2} presents comparisons between pristine AgInS$_{2}$ bulk
and AgInS$_{2}$ with Ag-In anti-stie disorder or Ag deficiency.
Different from CuInS$_{2}$, in AgInS$_{2}$, the Ag$^{+}$-In$^{3+}$ anti-site disorder-driven discontinuity
in the electrostatic potential is weakly screened by the S($p$)-Ag($d$) hybridization.
As shown in Fig.\ref{figure:DOS_Disorder_Defect_AgInS2}(a and b), the defect charges for the Ag and In sites
in the disordered atoms remain localized and un-screened, disconnected from the environment.
Due to this weak defect-environment hybridization, the electronic Coulomb scattering between
anti-site Ag($d$) electrons and VB electrons is weak.
This weaker scattering suggests that the coherent optical property in AgInS$_{2}$ is immune to
the anti-site disorder effects compared to the CuInS$_{2}$, explaining the restoration of the coherence
in the optical spectrum of Ag$_{1-x}$Cu$_{x}$In$_{1-y}$Ga$_{y}$S$_{2}$ quantum dot by reducing the Cu ratio $x$,
as shown in our experimental results in Figure~\ref{figure:PL_AIGS}.
Also, similar to CuInS$_{2}$, as shown in Fig.\ref{figure:DOS_Disorder_Defect_AgInS2}(a and c),
the Ag deficiency weakly impact the electronic structure compared to the pristine cases,
proving the immunity of the coherent optical spectrum in AgInS$_{2}$ quantum dot
concerning the structural and compositional inhomogeneity.

Table \ref{tab:U_values} presents on-site Hubbard $U$ parameters of TM($d$) orbital for CuInS$_{2}$ and AgInS$_{2}$,
comparing the pristine, the anti-site disorder, and the TM deficiency induced defect cases (See SM~\cite{suppl}),
consistent with electronic strutures in Fig.\ref{figure:DOS_Disorder_Defect_CuInS2} and Fig.\ref{figure:DOS_Disorder_Defect_AgInS2}.
The Cu($d$) $U$ parameter reveals relatively small changes in the consideration of the anti-site disorder and the TM defect cases,
implying the efficient screening of electrostatic defects induced by anti-site disorder and Cu deficiency,
consistent with Fig.\ref{figure:DOS_Disorder_Defect_CuInS2}.
On the other hand, for AgInS$_{2}$, the anti-site disorder induces a significant reduction of
$U$ value from 6.68 eV to 1.24 eV at the disordered site.
This significant variation of the Ag($d$) $U$ value
implies the weakly screened disorder-driven electrostatic defect, consistent with Fig.\ref{figure:DOS_Disorder_Defect_AgInS2}(a and b).
The Ag deficiency defects induce weak variations in the $U$ value for the neighboring Ag sites, similar to the case of CuInS$_{2}$,
consistent with Fig.\ref{figure:DOS_Disorder_Defect_AgInS2}(a and c).

Figure~\ref{figure:PL_AIGS}(a and b) presents the experimental
band edge photoluminescence spectra of Ag$_{1-x}$Cu$_{x}$In$_{1-y}$Ga$_{y}$S$_2$ as a function of Cu composition and its time-resolved decay.
The photoluminescence spectrum in Fig.\ref{figure:PL_AIGS}(a)
clearly reveals that with a small Cu doping, the lower energy side peak emerges, suggesting the formation of the
Cu atom at defect sites~\cite{liu2026core}.
By enhancing the Cu ratio, as shown in Fig.\ref{figure:PL_AIGS}(a),
the red shift is introduced for the higher energy main peak while the
lower energy side peak's energy position is fixed, consistent with Ref.~\cite{liu2026core}.
Notably, the lower energy side peak is broadened
by enhancing the Cu ratio and merges with the higher energy main peak spectra
when the Cu ratio is enhanced to 0.007, as shown in Fig.\ref{figure:PL_AIGS}(a).
The time-resolved photoluminescence decay results in Fig.\ref{figure:PL_AIGS}(b) shows that when the lower energy side peak starts to emerge by Cu doping, the lifetime of the spectrum decreases, and then by enhancing the Cu ratio further, the lifetime of the spectrum
increases, reaching the point where the Cu defect side peak merges with the higher energy main peak spectrum.

The Cu doping induced variation of the photoluminescence for Ag$_{1-x}$Cu$_{x}$In$_{1-y}$Ga$_{y}$S$_2$ quantum dot
in Fig.\ref{figure:PL_AIGS}(a and b) is a strong fingerprint of the present theoretical proposal on the role of $p$-$d$ hybridization for the coherent optical spectrum in this system, as demonstrated in Fig.\ref{figure:DOS_Disorder_Defect_CuInS2} and Fig.\ref{figure:DOS_Disorder_Defect_AgInS2}.
With a small Cu doping, the $p$-$d$ hybridization is enhanced for both Cu atoms at Ag sites and anti-site defect sites, as shown in Fig.\ref{figure:DOS_Disorder_Defect_CuInS2} and Fig.\ref{figure:DOS_Disorder_Defect_AgInS2}.
The Cu atoms at Ag sites introduce a strong repulsion between $p$-$d$ bonding and anti-bonding
thereby reducing the band gap, which is the source of the red shift of the higher energy main peak, as shown in Fig.\ref{figure:PL_AIGS}(a), supported by our computational result in Fig.\ref{figure:DOS_Disorder_Defect_CuInS2}(a) and Fig.\ref{figure:DOS_Disorder_Defect_AgInS2}(a).

On the other hand, the Cu atoms at anti-site defect sites
experiences a screening from the itinerant S($p$) electron as a quantum impurity problem~\cite{Anderson1961,Hewson1993,DoniachSondheimer},
as demonstrated in Fig.\ref{figure:DOS_Disorder_Defect_CuInS2}(b). This screening effect from the itinerant electron explains
the Cu doping induced broadening and the minor energy shift of the Cu defect-induced side peak
in Fig.\ref{figure:PL_AIGS}(a).
Furthermore, the computational results in Fig.\ref{figure:DOS_Disorder_Defect_CuInS2}(b)
imply that when the ratio of the Cu atoms at the anti-site defect site is substantial,
the screened Cu defect peak merges with the VBT, consistent with the
merging higher energy main peak and side peak at the Cu ratio of 0.007 in Fig.\ref{figure:PL_AIGS}(a).

These consistencies between (i) computational results of Fig.\ref{figure:DOS_Disorder_Defect_CuInS2} and Fig.\ref{figure:DOS_Disorder_Defect_AgInS2}, and (ii) experimental results in Fig.\ref{figure:PL_AIGS}, confirm the present theoretical prediction. Substantial $p$-$d$ hybridization induces an activation of the Cu($d$) Coulomb scattering channel by Cu doping.
Indeed, with a small Cu doping ratio in Ag$_{1-x}$Cu$_{x}$In$_{1-y}$Ga$_{y}$S$_2$
in Fig.\ref{figure:PL_AIGS}(b), merging of the higher energy main peak and the side peak
brings a restoration of the spectrum's lifetime due to the enhancement of screening
at Cu defect sites.
This activates randomly distributed Cu($d$) defect electron-environment electron hybridization, which induces Coulomb scattering-induced incoherence, driving full suppression of the band-edge spectrum when the Cu ratio is substantially enhanced, as shown in Ref.~\cite{jiang2023development}, reminiscent of Anderson localization~\cite{Anderson1958,Evers2008}.

In conclusion, using first-principles electronic structure calculations
on CuInS$_{2}$ and AgInS$_{2}$ bulk and experiments on the optical spectrum of Ag$_{1-x}$Cu$_{x}$In$_{1-y}$Ga$_{y}$S$_{2}$ quantum dot,
we demonstrate that the $p$-$d$ hybridization is a key factor
for the coherent optical properties of chalcopyrite semiconductors quantum dots.
The strong $p$-$d$ hybridization in the photo-induced charge carrier deteriorates the coherence of the optical properties of semiconductor quantum dots
from the Coulomb scattering of the open shell TM($d$) orbital.
This factor explains the Cu doping-induced incoherence of the optical spectrum of Ag$_{1-x}$Cu$_{x}$In$_{1-y}$Ga$_{y}$S$_{2}$.
These results provide a guideline for designing quantum dot materials immune to structural and compositional inhomogeneity,
which is unavoidable due to surface effects in quantum dots.

\acknowledgements
$Acknowledgements$

M.K. was supported by research funds for newly appointed professors of Jeonbuk National University in 2025,
the National Research Foundation of Korea(NRF) grant funded by the Korea government(MSIT) (RS-2026-25478055), the Korea Institute for Advanced Study (KIAS) individual Grants (No. CG083502), Korea Basic Science Institute (National Research Facilities and Equipment Center) Grant (No. RS-2024-00436672) funded by the Ministry of Education, Republic of Korea, and Korea Institute for Advancement of Technology (KIAT) grant funded by the Korea Government (MOTIE) (RS-2025-02214408, HRD Program for Industrial Innovation).
The DFT+$U$+$V$ calculation is supported by the Center for Advanced Computation at KIAS.

H.R, N.H and W.H was supported by research funds for newly appointed professors of Jeonbuk National University in 2024, Korea Basic Science Institute (National research Facilities and Equipment Center) grant funded by the Ministry of Education (grant No. RS-2024-00436672), and Global Learning \& Academic research institution for Master’s PhD students, and Postdocs (LAMP) Program of the National Research Foundation of Korea (NRF) grant funded by the Ministry of Education (No. RS-2024-00443714).
\bibliography{pd_hyb_ref}

\clearpage
\onecolumngrid
\section*{Supplemental Material}

\title {{\it Supplementary Material:}\\
  Role of $p$–$d$ Hybridization on Optical Properties of Chalcopyrite Semiconductors
}
\author{Neunghee Han}\thanks{These authors contributed equally to this work.}
\affiliation{Department of Semiconductor and Chemical engineering,
Jeonbuk National University, Jeonju 54896, Republic of Korea}
\author{Harang Kim}\thanks{These authors contributed equally to this work.}
\affiliation{Department of Semiconductor and Chemical engineering,
Jeonbuk National University, Jeonju 54896, Republic of Korea}
\author{Minjae Kim}\email{mjkim1985@jbnu.ac.kr}
\affiliation{Department of Semiconductor Science and Technology,
Jeonbuk National University, Jeonju 54896, Republic of Korea}
\author{Woonhyuk Baek}\email{whbaek@jbnu.ac.kr}
\affiliation{Department of Semiconductor and Chemical engineering,
Jeonbuk National University, Jeonju 54896, Republic of Korea}
\date{\today}
\maketitle

\section{Computational Details}

We performed first-principles calculation
using self-consistent extended Hubbard interactions called as density functional theory plus $U$ plus $V$ (DFT+$U$+$V$) method\cite{PhysRevX.5.011006,lee2020first,tancogne2020parameter,jang2023intersite,hohenberg1964inhomogeneous,kohn1965self,leiria2010extended} implemented in QUANTUM ESPRESSO package~\cite{giannozzi2009quantum}.
Using a newly developed extended ACBN0 ($e$ACBN0) method\cite{PhysRevX.5.011006,lee2020first,tancogne2020parameter,jang2023intersite},
we can obtain on-site Hubbard interactions $U$ of atomic orbitals and $V$ between them self-consistently.
For DFT+$U$+$V$ calculations, we use norm-conserving pseudopotentials provided by the PseudoDojo project for atomic potential, generalized gradient approximation (GGA)
for exchange-energy functionals for DFT calculations.
We adopt experimental crystal structures of CuInS$_{2}$ and AgInS$_{2}$
for the calculations of electronic structures.\cite{Spiess1974,Delgado2001}
A $21 \times 21 \times 11$ $k$-point mesh and a kinetic-energy cutoff of 100 Ry were used for the pristine unit cell,
whereas a $4 \times 4 \times 4$ mesh and an 80 Ry cutoff were adopted for the $2 \times 2 \times 1$ supercell calculations.
The Hubbard $U$ and $V$ parameters were self-consistently determined for each material and supercell configuration using the $e$ACBN0 method\cite{PhysRevX.5.011006,lee2020first,tancogne2020parameter,jang2023intersite}.

\section{Experimental Method}

Oleylamine (OLA, $>$50\%, TCI), 1-Octadecene (ODE, 90.0\%, Rhawn) ,Silver(I) Iodide (AgI, $>$99.0\%, TCI), Indium(III) iodide(InI$_{3}$,  99.999\%, Thermo Scientific), Gallium(III) iodide (GaI$_{3}$, 99.99\%, Sigma-Aldrich), Copper(I) iodide, (CuI, 98.0\%, SAMCHUN), Sulfur powder (S, 99\%, DAEJUNG), 1-dodecanethiol (DDT, 98\%, DAEJUNG), Tri-n-octylphosphine (TOP, $>$85\%, TCI)

To synthesize Cu-doped AIGS QDs, 0.2 mmol of AgI, 0.2 mmol of InI$_{3}$, and 0.7 mmol of GaI$_{3}$ were added to a vial containing oleylamine (OLA, 5 mL) and 1-octadecene (ODE, 5 mL), followed by addition of a 42 mM CuI–OLA solution at the desired ratio. The reaction mixture was degassed at 120~$^\circ$C for 30 min. Meanwhile, a separate vial containing 1 M S–OLA solution was prepared at 100~$^\circ$C. After completion of degassing, the atmosphere was switched to N$_{2}$, 1.6 mL of 1-dodecanethiol (DDT) was rapidly injected, and the mixture was allowed to react for 5 min. Subsequently, 1.3 mL of the S–OLA solution was injected, the temperature was raised to 280~$^\circ$C, and the reaction was maintained for an additional 5 min. After the reaction was complete, the vial was rapidly cooled to 180~$^\circ$C, followed by injection of trioctylphosphine (TOP, 4 mL) and further stirring for 20 min for surface treatment. Finally, the heating mantle was removed, and the reaction mixture was allowed to cool naturally to room temperature.

Photoluminescence (PL) spectroscopy were obtained using a Horiba (FluoroMax Plus) spectrophotometer equipped with a Photomultiplier R928P base detector installed at the Center for University Wide Research Facilities (CURF) at Jeonbuk National University.
Time-Resolved Photoluminescence spectra (Horiba, Fluorolog-QM) and Contact angles (Ossila, L2004A1) were measured at the Future Energy Convergence Core Center (FECC).

\end{document}